
\input phyzzx
\vsize 9.25in
\hsize 6.3in
\def\cMq{{\cal M}_q}
\def\bcMq{\bar {\cal M}_q}
\def\bq{\bar q}
\def\bphi{\bar \phi}
\def\bt{\bar t}
\def\hJ{\hat J}
\def\hP{\hat P}

\rightline{UATP-93/06}
\rightline{December 1993}
\vskip 0.2in
\centerline{\seventeenbf Wilson Loops and Black Holes}
\centerline{\seventeenbf in 2+1 dimensions}
\vskip 0.75in
\centerline{\caps Cenalo Vaz\footnote{\dagger}{Internet:
cvaz@mozart.si.ualg.pt}}
\centerline{\it Unidade de Ci\^encias Exactas e Humanas}
\centerline{\it Universidade do Algarve}
\centerline{\it Campus de Gambelas, P-8000 Faro, Portugal}
\vskip 0.2in
\centerline{and}
\vskip 0.2in
\centerline{\caps Louis Witten\footnote{\dagger\dagger}{Internet:
witten@ucbeh.san.uc.edu}}
\centerline{\it Department of Physics}
\centerline{\it University of Cincinnati}
\centerline{\it Cincinnati, OH 45221-0011, U.S.A.}
\vskip 0.75in
\centerline{\bf \caps Abstract}
\vskip 0.2in

In 2+1 dimensional Chern-Simons gravity, Wilson loops in the three
dimensional Anti de Sitter group, $SO(2,2)$, reproduce the spinning
black hole of Ba\~nados, Teitelboim and Zanelli (BTZ) by naturally
duplicating the necessary identification of points of a four
dimensional globally $SO(2,2)$ invariant space in which the hole
appears as an embedding.
\vfill
\eject
Pure gravity, Anti-de Sitter (AdS) and de Sitter (dS) gravity and
supergravity in three dimensions can be formulated as a
Chern-Simons gauge theories of Poincar\'e group, $ISO(2,1)$, the
Anti-de Sitter group, $SO(2,2)$, the de Sitter group, $SO(3,1)$,
and their various supersymmetric extensions.${}^{1,2}$  The
Chern-Simons action is topological and in all cases the theory is
finite and solvable both on the classical and the quantum
levels${}^3$.  Moreover, the precise relationship between the
Poincar\'e Chern-Simons theory and the spacetime of Einstein's
gravity in three dimensions has been well established.${}^4$

Recently Ba\~nados, Teitelboim and Zanelli${}^{5,6}$ (BTZ)
discovered that three dimensional AdS gravity admits a spinning
black hole solution which arises by identification of points of AdS
space by a discrete subgroup of $SO(2,2)$.  In Einstein gravity,
these identifications must be made by hand and one is led to seek
a more natural framework for them.  The Chern-Simons approach
provides precisely such a framework.  It is well known, for
example, that the multiconical structure of spacetime in pure
(Poincar\'e) gravity arises from multivalued gauge transformations
on the coordinates of a globally trivial Minkowski spacetime.${}^7$
The multivaluedness of the gauge transformations leads to conical
identifications of points in Minkowski space.  Our aim in this
paper is to provide a similar description of the BTZ black hole.

In ref.\ (2) gravity in three dimensions was interpreted as a gauge
theory with a Chern-Simons action
$$\eqalign{I_{C.S.}~~ &=~~ -~{1 \over 2}~ {\rm Tr}~ \int_M~ A
\wedge (dA~ +~ {2 \over 3}~ A \wedge A) \cr &=~~ -~ {1 \over 2}~
\int_M~ \gamma_{BC} A^B \wedge (dA^C~ +~ {1 \over 3}~ f^C_{DE}  A^D
\wedge A^E), \cr} \eqno(1)$$
where $A_\mu$ is the gauge connection, $f^C_{DE}$ are the structure
constants of the gravity group, $X_A$ are the generators of its Lie
algebra, $\gamma_{AB} = 2 {\rm Tr} (X_A X_B)$ plays the role of a
metric on the Lie algebra and $M$ is an arbitrary three manifold.
Consider the basis
$$[\hJ_{ab},\hJ_{cd}]~~ =~~ \eta_{ad} \hJ_{bc}~ +~ \eta_{bc}
\hJ_{ad}~ -~ \eta_{ac} \hJ_{bd}~ -~ \eta_{bd} \hJ_{ac} \eqno(2)$$
of the Lie algebra $so(2,2)$ of $SO(2,2)$.  The roman indices
$a...~ \epsilon~ \{0,1,2,3\}$, and $\eta_{ab}$ is the metric of a
four dimensional flat space, which we denote by $\cMq$, with
signature $\eta_{ab} = {\rm diag} (-,-,+,+)$.  $SO(2,2)$ admits two
Casimir operators, which in this basis are $J_{ab}J^{ab}$ and
$\epsilon_{abcd} J^{ab} J^{cd}$.  Expanding $A_\mu$ according to
$$A_\mu~~ =~~ {1 \over 2}~ \omega^{ab}_\mu \hJ_{ab}, \eqno(3)$$
the action in (1) can be put in the form
$$I_{C.S.}~~ =~~ -~ \int_M \epsilon_{abcd}~ \omega^{ab} \wedge ( d
\omega^{cd}~ +~ {1 \over 6}~ f^{cd}_{(ab)(ef)} \omega^{ab} \wedge
\omega^{ef}), \eqno(4)$$
where
$$f^{cd}_{(ab)(ef)}~~ =~~ \eta_{af} \delta^c_b \delta^d_e~ +~
\eta_{be} \delta^c_a \delta^d_f~ -~ \eta_{ae} \delta^c_b
\delta^d_f~ -~ \eta_{bf} \delta^c_a \delta^d_e \eqno(5)$$
and $\gamma_{AB} = \epsilon_{abcd}$, $\epsilon^{0123} = + 1$
serves as the metric on $so(2,2)$. The metric has been chosen so
that (4) has a non-degenerate Poincar\'e limit.  In a time-space
decomposition, the action in (4) can also be written in the form
$$I_{C.S.}~~ =~~ \int dt \int_\Sigma \epsilon_{abcd} \epsilon^{ij}
\left( \omega^{ab}_i \partial_t \omega^{cd}_j~ -~ \omega^{ab}_0
F^{cd}_{ij} \right) \eqno(6)$$
where $i,j~ \epsilon~ \{1,2\}$ are spatial indices in $M$ and
$$F^{ab}_{\mu\nu}~~ =~~ \partial_\mu \omega^{ab}_\nu~ -~
\partial_\nu \omega^{ab}_\mu~ +~ \left( \omega^a_{f\mu}
\omega^{fb}_\nu~ -~ \omega^a_{f\nu} \omega^{fb}_\mu \right)
\eqno(7)$$
is the $SO(2,2)$ curvature tensor.  From (6) follow the free field
Poisson brackets,
$$\{\omega^{ab}_i(x), \omega^{cd}_j(y)\}_{P.B.}~~ =~~ -~
\epsilon^{abcd} \epsilon_{ij} \delta^{(2)}(x,y) \eqno(8)$$
and, as $\omega^{ab}_0$ is a Lagrange multiplier, the constraints
$$\epsilon^{ij} F^{ab}_{ij}~~ =~~ 0. \eqno(9)$$

An alternative formulation can be given in the basis $\hP_a,~
\hJ_a$ of $so(2,2)$ defined by
$$[\hP_a, \hP_b]~ =~ \lambda \epsilon_{abc} J^c,~~~~ [\hP_a,
\hJ_b]~ =~ \epsilon_{abc} P^c,~~~~ [\hJ_a, \hJ_b]~ =~
\epsilon_{abc} J^c, \eqno(10)$$
where $\lambda$ is the ``cosmological constant'' and the roman
indices $a...~ \epsilon~ \{0,1,2\}$ are raised and lowered by
$\eta_{ab} = {\rm diag} (-,+,+)$, if $A_\mu$ is expanded according
to
$$A_\mu = e^a_\mu {\hat P}_a + \omega^a_\mu {\hat J_a}\eqno(11)$$
where $e^a_\mu$ is the dreibein and $\omega^a_\mu$ is the spin
connection.  Of the two quadratic Casimirs, $\hP \cdot \hJ$ and
$\hP^2 + \hJ^2 / \lambda^2$, the former must be used if one is
interested in a non-degenerate invariant bilinear form in the
Poincar\'e limit.  The Chern-Simons action may then be cast into
the form
$$I_{C.S}~~ =~~ \int dt \int_\Sigma \eta_{ab} \epsilon^{ij} \left[
e^a_i \partial_t \omega^b_j~ -~ e^a_0 F^b_{ij} [\omega]~ -~
\omega^a_0 F^b_{ij} [e] \right], \eqno(12)$$
where $F^a_{\mu\nu}[e]$ and $F^a_{\mu\nu}[\omega]$ are the torsion
and curvature respectively.

The two descriptions of AdS gravity may be connected by the
following correspondence between the components of the connection
in the two bases:
$$\eqalign{\omega^{01}_\mu~ &=~ {\sqrt{\lambda}} e^0_\mu,~~~~~~~~~~
\omega^{23}_\mu~ =~ {\sqrt{\lambda}} \omega^0_\mu,\cr
\omega^{12}_\mu~ &=~ - {\sqrt{\lambda}} e^1_\mu,~~~~~~~~
\omega^{03}_\mu~ =~ {\sqrt{\lambda}} \omega^1_\mu,\cr
\omega^{13}_\mu~ &=~ {\sqrt{\lambda}} e^2_\mu,~~~~~~~~~~
\omega^{02}_\mu~ =~ {\sqrt{\lambda}} \omega^2_\mu,\cr}, \eqno(13)$$
The authors in ref.\ (8) compute the group element associated with
the BTZ solution in the second basis.  We will use the Chern-Simons
theory to reproduce the identifications required to construct the
black hole as an embedding, so we will work in the first.

The generators $\hJ_{ab}$ admit the following irreducible infinite
dimensional representation on $\cMq$,
$$\hJ_{ab}~~ =~~ q_a \partial_b~ -~ q_b \partial_a ,\eqno(14)$$
where $q^a$ are coordinates in $\cMq$.  Following Witten's${}^2$
suggestion, if $p_a$ is conjugate to $q^a$, spinless point sources
can be coupled to the field action above by adding to (4) the
Borel-Weil-Bott action
$$I_S~~ =~~ \int dt~ \left( p_a D_t q^a~ +~ {\rm constraints}
\right), \eqno(15)$$
in which $D_t$ is the $SO(2,2)$ covariant derivative,
$$D_t~~ =~~ \partial_t~~ +~~ {1 \over 2}~ t^\mu~ \omega^{mn}_\mu~
\hJ_{mn}, \eqno(16)$$
where $t^\mu$ is tangent to the trajectory of the source in $M$.
The $q^a(t)$ represent points in $\cMq$.  They are functions of
time, mapping points on the trajectory of the particle in $M$ to
points in $\cMq$ so that
$$q^a(t)~~ =~~ q^a[x^\mu(t)] \eqno(17)$$
is a functional of points $P[x^\mu (t)]$ in $M$. A source at the
origin of $M$,  will therefore modify the constraints in (9) to
read
$$\epsilon^{ij} F^{ab}_{ij}~~ =~~ (p^a q^b~ -~ p^b q^a) \delta^2
(x)~~ =~~ S^{ab} \delta^2 (x). \eqno(18)$$
where the $S^{ab}$ are conserved charges, by Noether's theorem.
The source action has an obvious generalization to the case of $N$
non-interacting particles.

The $(p_a, q^a)$, cannot be identified with the spacetime momenta
and coordinates of the particle.  They transform as $SO(2,2)$
vectors.  Consider now the coordinates $\bq(P)$, constructed from
the coordinates $q(P)$ by the following gauge transformation
$$\bq(P)~~ =~~ g(P) q(P)~~ =~~ {\cal P} \exp (\int_*^P dx^\mu
A_\mu)~ q(P), \eqno(19)$$
where ${\cal P}$ represents path ordering.  The gauge
transformation
$$g(P)~~ =~~  W_{*P}~~ =~~ {\cal P} \exp (\int_*^P dx^\mu A_\mu)
\eqno(20)$$
is the Wilson line from the origin, $*$ in $M$, to the point $P$.
We think of the $\bq(P)$ as living in an internal space, donoted
hereafter by $\bcMq$. These coordinates satisfy unusual boundary
conditions when $S^{ab} \neq 0$.

To see this consider a single source located at $*$, and a
trajectory $\gamma$ which loops about the source.  Let the point
$P$ on $\gamma$ be mapped to the point $q(P)$ in $\cMq$, so that
$\bq(P)$ in $\bcMq$ is given by (19).  Traversing the loop $\gamma$
once in the counterclockwise direction, and returning to $P$, we
find that the same point in $\bcMq$ is also given by
$$\bq(P)~~ =~~ W_{*P} W_\gamma q(P)~~ =~~ W_{*P} W_\gamma W_{P*}
\bq(P) \eqno(21)$$
where we have used $W_{*P} W_{P*} = 1$ above.  If $W_*$ is the
infinitesimal loop about the origin,
$$W_\gamma~~ =~~ W_{P*} W_* W_{*P} \eqno(22)$$
by a deformation of the loop $\gamma$.  Thus the point in $\bcMq$
associated with $\bq(P)$ must be identified with the point $W_*
\bq(P)$.  For non-vanishing $S^{ab}$ this is a non-trivial matching
condition because, by the non-abelian Stokes theorem,
$$W_*~~ =~~ e^{S^{ab} \hJ_{ab}}. \eqno(23)$$
If there are many sources, located at the points $P_1,P_2,...P_N$,
the matching condition takes the form
$$\bq(P)~~ =~~ {\bar C}_1 {\bar C}_2~ \cdot~ \cdot~ \cdot~ {\bar
C}_K~ \bq(P) \eqno(24)$$
where each $C_I$ is an infinitesimal circle about the point $P_I$,
and ${\bar C}_I$ is the result of parallel transporting the source
to the origin, i.e.,
$${\bar C}_I~~ =~~ W_{*I} C_I W_{I*}~~ =~~ e^{{\bar S}^{ab}
\hJ_{ab}} \eqno(25)$$
where
$${\bar S}^{ab} \hJ_{ab}~~ =~~ W_{*I} S^{ab} \hJ_{ab} W_{I*}
\eqno(26)$$
and only those sources lying within the loop $\gamma$ are included
in the product on the right hand side of (24).  It is necessary of
course to fix an ordering of the sources, which can be done, for
example, by increasing azimuthal angle in $M$.  The formal matching
conditions in (24) are difficult to describe geometrically.
Note, however, that the $\bq(P)$ defined by (19) are sensitive only
to gauge transformations at $*$.  This is because, under an
arbitrary gauge transformation
$$\eqalign{W_{*P}~~ &\rightarrow~~ U(*) W_{*P} U^{-1} (P) \cr
q(P)~~ &\rightarrow~~ U(P) q(P) \cr} \eqno(27)$$
giving
$$\bq(P)~~  \rightarrow U(*) \bq (P). \eqno(28)$$
where we have used (19).  Note also that the derivation of these
matching conditions is independent of the nature of $M$, therefore
the structure of $\bcMq$ is also independent of it.

The transformation (19) can be used to reconstruct spacetime.  It
has already been shown to reproduce the multiconical structure of
pure gravity${}^7$.  In AdS gravity it is responsible for the
appearance of a massive, spinning black hole as an embedding, not
in $\cMq$ but in $\bcMq$.  AdS space may be thought of as an
embedding in $\cMq$ through the constraint
$$-(q^0)^2~~ -~~ (q^1)^2~~ +~~ (q^2)^2~~ +~~ (q^3)^2~~ =~~ -~ l^2
\eqno(29)$$
How one chooses to parametrize the constraint then determines which
part of the manifold one covers. The conventional parametrization
covers the entire manifold.  Let
$$\eqalign{&(q^2)^2~ +~ (q^3)^2~ =~ r_q^2,\cr &q^2~ =~ r_q \cos
\phi_q,~~~~ q^3~ =~ r_q \sin \phi_q\cr &(q^0)^2~ +~ (q^1)^2~ =~
r_q^2~ +~ l^2,\cr &q^0~ =~ {\sqrt{r_q^2 + l^2}} \cos(t_q/l),~~~~
q^1~ =~ {\sqrt{r_q^2 + l^2}} \sin(t_q/l) \cr} \eqno(31)$$
where $r_q~ \epsilon~ [0,\infty)$ and $t_q/l, \phi~ \epsilon~
[0,2\pi)$ are periodic coordinates.  The metric $ds^2 = - \eta_{ab}
dq^a dq^b$ is then
$$ds^2~~ =~~ ({{r_q^2} \over {l^2}} + 1) dt_q^2~~ -~~ {{dr_q^2}
\over {({{r_q^2} \over {l^2}} + 1)}}~~ -~~ r_q^2 d\phi_q^2,
\eqno(31)$$
but $t_q$ is still a periodic coordinate with period $2 \pi l$.
This means that there are closed time-like curves.  To avoid them
one no longer identifies $t_q = 0$ with $t_q = 2 \pi l$, and
arrives at the usual AdS space which is therefore the universal
covering of the space in (31).  In an alternative parametrization,
let
$$\eqalign{&-~ (q^0)^2~~ +~~ (q^3)^2~~ =~~ -~ r_q^2\cr &(q^0)~~ =~~
r_q \cosh \phi_q,~~~~~ (q^3)~~ =~~ r_q \sinh \phi_q\cr &-~
(q^1)^2~~ +~~ (q^2)^2~~ =~~ r_q^2~~ -~~ l^2\cr &q^1~~ =~~
{\sqrt{r_q^2~ -~ l^2}} \sinh(t_q/l),~~~~~ q^2~~ =~~ {\sqrt{r_q^2~ -
{}~ l^2}} \cosh(t_q/l)~~~~ r_q~ >~ l \cr &q^1~~ =~~ {\sqrt{l^2~ -~
r_q^2}} \cosh(t_q/l),~~~~~ q^2~~ =~~ {\sqrt{l^2~ -~ r_q^2}}
\sinh(t_q/l)~~~~ r_q~ <~ l \cr} \eqno(32)$$
where $r_q,\phi_q$ and $t_q$ are all non-compact coordinates.  The
metric $ds^2 = - \eta_{ab} dq^a dq^b$ is then
$$ds^2~~ =~~ ({{r_q^2} \over {l^2}} - 1) dt_q^2~~ -~~ {{dr_q^2}
\over {({{r_q^2} \over {l^2}} - 1)}}~~ -~~ r_q^2 d\phi_q^2,
\eqno(33)$$
but $\phi_q$ is non compact, and (33) describes the universal
covering of AdS space. Imposing periodicity on $\phi$ by hand,
i.e., identifying $\phi_q = 0$ and $\phi_q = 2\pi$, we recover the
AdS black hole spacetime of ref.\ (6) with ``mass'' $M = 1$ and
zero angular momentum.

To include angular momentum and arbitrary mass, consider a single
source situated at the origin of $M$, satisfying
$$S^{03}~~ =~~ 2\pi A,~~~~~ S^{12}~~ =~~ 2 \pi B \eqno(34)$$
and all others zero.  In a polar coordinatization of $\Sigma$, the
solution of the constraint equations reads
$$\omega^{03}_\phi~~ =~~ A,~~~~~ \omega^{12}_\phi~~ =~~ B
\eqno(35)$$
and all other spatial components vanish.  The time components of
the connection are Lagrange multipliers which may be selected
arbitrarily.  To recover the BTZ black hole we must also pick
$$\omega^{03}_t~~ =~~ S~~ \neq~~ 0. \eqno(37)$$
With this choice of connection, the coordinates $\bq(P)$ defined in
(19) become
$$\bq(P)~~ =~~ e^{\left(-(\int^t dt \omega^{o3}_t~ +~ \int^\phi
d\phi \omega^{03}_\phi) {\hat J}_{03}~ -~ \int^\phi d\phi
\omega^{12}_\phi {\hat J}_{12} \right)}~ q(P) \eqno(38)$$
where $\phi$ and $t$, which appear in the exponent above are the
angular coordinate and time coordinate respectively in $M$.
Equation (38) gives for the barred coordinates
$$\eqalign{\bq^0~~ &=~~ \cosh(St + A\phi)q^0~~ -~~ \sinh
(St+A\phi)q^3 \cr \bq^1~~ &=~~ \cosh(B\phi)q^1~~ -~~ \sinh
(B\phi)q^2 \cr \bq^2~~ &=~~ \cosh(B\phi)q^2~~ -~~ \sinh (B\phi)q^1
\cr \bq^3~~ &=~~ \cosh(St+A\phi)q^3~~ -~~ \sinh (St+A\phi)q^0 \cr
} \eqno(39)$$
Once again, imposing the AdS constraint with the same
parametrization as in (32) but on the barred coordinates one finds
$$\eqalign{&-~ (\bq^0)^2~~ +~~ (\bq^3)^2~~ =~~ -~ r_q^2\cr
&(\bq^0)~~ =~~ r_q \cosh \bphi_q,~~~~~ (\bq^3)~~ =~~ r_q \sinh
\bphi_q\cr &-~ (\bq^1)^2~~ +~~ (\bq^2)^2~~ =~~ r_q^2~~ -~~ l^2\cr
&\bq^1~~ =~~ {\sqrt{r_q^2~ -~ l^2}} \sinh(\bt_q/l),~~~~~ \bq^2~~
=~~ {\sqrt{r_q^2~ -~ l^2}} \cosh(\bt_q/l)~~~~ r_q~ >~ l \cr
&\bq^1~~ =~~ {\sqrt{l^2~ -~ r_q^2}} \cosh(\bt_q/l),~~~~~ \bq^2~~
=~~ {\sqrt{ l^2~ -~ r_q^2}} \sinh(\bt_q/l)~~~~ r_q~ <~ l. \cr}
\eqno(40)$$
The barred coordinates $\bt_q$ and $\bphi_q$ are related to the
previous ones in (32) by
$$\bt_q~~ =~~ t_q~ -~ l B\phi,~~~~~~ \bphi_q~~ =~~ \phi_q~ -~ St~ -
{}~ A\phi \eqno(41)$$
where $\phi$ and $t$ are coordinates in $M$ and the metric $ds^2 =
-\eta_{ab} d\bq^a d\bq^b$ is evidently
$$ds^2~~ =~~ ({{r_q^2} \over {l^2}} - 1) d\bt_q^2~~ -~~ {{dr_q^2}
\over {({{r_q^2} \over {l^2}} - 1)}}~~ -~~ r_q^2 d\bphi_q^2,
\eqno(42)$$
in terms of the barred coordinates.

When $|B| < |1-A|$, an identification $t \equiv t_q$ and $\phi
\equiv \phi_q$ gives the BTZ black hole.  Let
$$B~ =~ {{r_-} \over l},~~~~~ (1-A)~ =~ {{r_+} \over l},~~~~~ S~ =~
{{r_-} \over {lr_+}}, \eqno(43)$$
where $r_- < r_+$.  Indeed, $r_-$ and $r_+$ are the inner and outer
horizons of the AdS black hole,
$$r_\pm^2~~ =~~ {{l^2} \over 2}~ \left( M~~ \pm~~ {\sqrt{M^2~ -~
(J/l)^2}} \right), \eqno(44)$$
where $M$ and $J$ are its mass and angular momentum, and the metric
$$ds^2~~ =~~ ({{r_q^2} \over {l^2}} - 1) (dt_q~ -~ r_- d\phi_q
)^2~~ -~~ {{dr_q^2} \over {({{r_q^2} \over {l^2}} - 1)}}~~ -~~
r_q^2 ({{r_+} \over l}d\phi_q~ -~{{r_-} \over {lr_+}} dt)^2,
\eqno(45)$$
can be brought into the standard (BTZ) form
$$ds^2~~ =~~ ({{r^2} \over {l^2}} - M + {{J^2} \over {4r^2}}) dt^2~
-~ {{dr^2} \over {({{r^2} \over {l^2}} - M + {{J^2} \over
{4r^2}})}}~ -~ r^2 (d\phi~ -~ {J \over {2r^2}} dt)^2 \eqno(46)$$
by the coordinate transformation
$$\eqalign{r^2~~ &=~~ \left[ {{r_+^2~ -~ r_-^2} \over {l^2}}
\right] r_q^2~~ +~~ r_-^2 \cr t~~ &=~~ {l \over {r_+}}~ t_q \cr
\phi~~ &=~~ \phi_q. \cr} \eqno(47)$$
For $r_q > l$ (46) gives the region $r > r_+$ and for $r_q < l$ it
gives the region $r_- < r < r_+$.

To recover the metric in the region $r < r_-$, we consider the
parametrization
$$\eqalign{&-~ (q^0)^2~~ +~~ (q^3)^2~~ =~~ r_q^2\cr &(q^0)~~ =~~
r_q \sinh \phi_q,~~~~~ (q^3)~~ =~~ r_q \cosh \phi_q\cr &(q^1)^2~~ -
{}~~ (q^2)^2~~ =~~ r_q^2~~ +~~ l^2\cr &q^1~~ =~~ {\sqrt{r_q^2~ +~
l^2}} \cosh(t_q/l),~~~~~ q^2~~ =~~ {\sqrt{r_q^2~ +~ l^2}}
\sinh(t_q/l)} \eqno(48)$$
which gives
$$ds^2~~ =~~ -~ ({{r_q^2} \over {l^2}} + 1) dt_q^2~~ -~~ {{dr_q^2}
\over {({{r_q^2} \over {l^2}} + 1)}}~~ +~~ r_q^2 d\phi_q^2.
\eqno(49)$$
When this parametrization is applied to $\bcMq$ as we did in the
previous two cases, the resulting metric can be cast into the form
$$ds^2~~ =~~ -~ ({{r_q^2} \over {l^2}} + 1) (dt_q~ -~ r_- d\phi_q
)^2~~ -~~ {{dr_q^2} \over {({{r_q^2} \over {l^2}} + 1)}}~~ +~~
r_q^2 ({{r_+} \over l}d\phi_q~ -~{{r_-} \over {lr_+}} dt)^2,
\eqno(50)$$
which yields the metric (46) in the region $0 < r < r_-$ after
making the coordinate transformation (47) with $r_q^2 \rightarrow -
 r_q^2$. The transformation is valid only in the region $ 0 \leq
r_q^2 < l^2 r_-^2 / (r_+^2 - r_-^2)$.

We have shown that the identifications necessary to construct the
BTZ black hole arise naturally from the Chern-Simons approach to
gravity in three dimensions.  Our construction may be generalized
to supergravity, which has a similar description as a Chern-Simons
theory.${}^{2,9}$  The advantage of a Chern-Simons description is
also the ease with which quantum effects may be described.${}^4$
These and other issues of black hole physics in three dimensions
will be discussed more fully elsewhere.
\vskip 0.5in

\noindent{\bf Acknowledgement}

\noindent This work was supported in part by  NATO under contract
number CRG 920096. L.W. acknowledges the partial support of the U.
S. Department of Energy under contract number DOE-FG02-84ER40153.
\vskip 0.5in

\noindent{\bf References}

{\item{1.}}S Deser, R. Jackiw and G. 't Hooft, Ann Phys. {\bf 152}
(1984) 220; S. Deser and R. Jackiw, {\it ibid} {\bf 153} (1984)
405.

{\item{2.}}A. Achucharro and P. K. Townsend, Phys. Letts. {\bf
B180} (1986) 89; E. Witten, Nucl. Phys. {\bf B311} (1988) 46; K.
Koehler, F. Mansouri, C. Vaz and L. Witten, Mod. Phys. Letts. {\bf
A5} (1990) 935; J. Math. Phys. {\bf 32}, (1990) 239.

{\item{3.}}E. Witten, Nucl. Phys. {\bf B323} (1989) 113; O. F.
Dayi, Phys. Letts. {\bf B234} (1990) 25.

{\item{4.}}K. Koehler, F. Mansouri, C. Vaz and L. Witten, Nucl.
Phys. {\bf B348} (1991) 373; {\it ibid} {\bf B358} (1991) 677.

{\item{5.}}M. Ba\~nados, C. Teitelboim and J. Zanelli, Phys. Rev.
Letts. {\bf 69} (1992) 1894.

{\item{6.}}M. Ba\~nados, C. Teitelboim and J. Zanelli, Phys. Rev.
{\bf D48} (1993) 1506.

{\item{7.}}C. Vaz and L. Witten, Nucl. Phys. {\bf B368} (1992) 509;
in {\it Proc. of the XX International Conference on Differential
Geometric Methods in Theoretical Physics}, Vol. II, eds. Sultan
Catto and Alvany Rocha, World Scientific, N.Y. 1992.

{\item{8.}}D. Cangemi, M. Leblanc and R. B. Mann, Phys. Rev. {\bf
D48} (1993) 3606.

{\item{9.}}Olivier Coussarert and Marc Henneaux, Phys. Rev. Letts.
{\bf 72} (1994) 183.
\end